# Managing Industrial Communication Delays with Software-Defined Networking

Rutvij H. Jhaveri[*], Rui Tan[†], Arvind Easwaran[†] and Sagar V. Ramani[‡]
[*] Delta – NTU Corporate Laboratory for Cyber-Physical Systems
School of Electrical and Electronic Engineering, Nanyang Technological University, Singapore
Email: rhjhaveri@ntu.edu.sg
[†]School of Computer Science and Engineering, Nanyang Technological University, Singapore
[‡]Department of Computer Engineering, Government Polytechnic, Porbandar, India

*Abstract*—Recent technological advances have fostered the development of complex industrial cyber-physical systems which demand real-time communication with delay guarantees. The consequences of delay requirement violation in such systems may become increasingly severe. In this paper, we propose a contract-based fault-resilient methodology which aims at managing the communication delays of real-time flows in industries. With this objective, we present a light-weight mechanism to estimate end-to-end delay in the network in which the clocks of the switches are not synchronized. The mechanism aims at providing high level of accuracy with lower communication overhead. We then propose a contract-based framework using software-defined networking (SDN) where the components are associated with delay contracts and a resilience manager. The proposed resilience management framework contains: (1) contracts which state guarantees about components' behaviors, (2) observers which are responsible to detect contract failure (fault), (3) monitors to detect events such as run-time changes in the delay requirements and link failure, (4) control logic to take suitable decisions based on the type of the fault, (5) resilience manager to decide response strategies containing the best course of action as per the control logic decision. Finally, we present a delay-aware path finding algorithm which is used to route/reroute the real-time flows to provide resiliency in the case of faults and, to adapt to the changes in the network state. Performance of the proposed framework is evaluated with the Ryu SDN controller and Mininet network emulator.

## I. INTRODUCTION

Industrial cyber-physical systems (ICPS) represent significant innovations in the development of information and communication technologies. They have enormous potential to improve smart manufacturing systems in the fourth industrial revolution (Industry 4.0) [1]. They are the systems incorporating sensors, actuators and computational entities which constantly interact in real time with the surrounding physical industrial components. In order to operate such systems more efficiently, it is important to understand the interaction between the cyber and the physical entities [2]. Network resiliency becomes a key aspect of such systems in order to provide fault-tolerant real-time communication even when challenged by network failures or dynamically changing needs of flows.

Time-critical applications in ICPS require guaranteed upper bounds on end-to-end delay for timely delivery of data packets between network hosts [3]. Thus, it becomes a key problem to satisfy the delay requirements of the real-time flows by providing resiliency to the faults in an efficient way. Traditional fault-resilient approaches for managing communication delays for such systems need the use of costly hardware or software components. Recently software-defined networking (SDN) has become increasingly popular as it separates the data plane from the control plane where a centralized SDN controller contains all the control functionalities (refer Figure 1). With the global view and network intelligence, an SDN controller can significantly reduce the management overheads of the traditional approaches [4]. At the same time, it has the potential to provide fault-resiliency in the network.

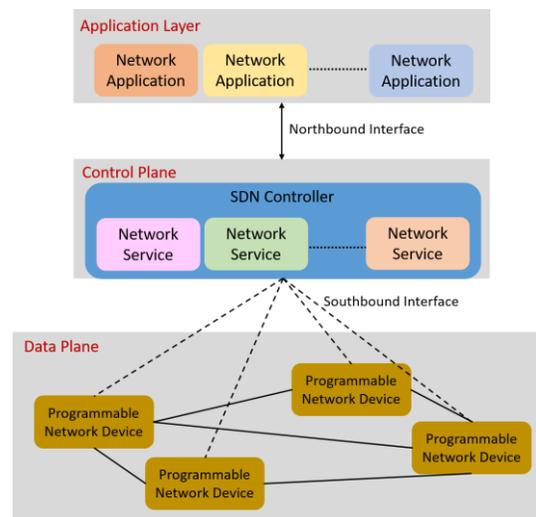

Fig. 1. Software-defined networking architecture

In this paper, therefore, we propose a fault-resilient SDN-based framework which aims at managing end-to-end communication delays for real-time flows in ICPS. Meanwhile, selection of the model to estimate end-to-end delay is imperative for time-critical industrial operations. As the traditional approaches have complex and expensive methodology to estimate end-to-end delay, we present a light-weight end-to-end delay estimation mechanism with software-defined networking which utilizes default probe packets in order to limit the communication overhead. As our motivation to adopt this mechanism to estimate end-to-end delay in the fault-resilient framework, we evaluate the accuracy of this mechanism in distinct network conditions. Going further, a contract-based methodology [1] is integrated with the proposed SDN-based framework where the component (SDN controller) is associated with contracts and a resilience manager (RM). This methodology allows the system to detect faults (contract violation) and, react to the fault in an efficient way. When the RM is unable to provide a feasible solution to a specific fault, it issues a warning and notifies the corresponding software component. In order to evaluate the performance of our approach, we carry out experiments with Ryu SDN controller and Mininet emulator with UDP traffic. The results show that our approach outperforms other approaches and improves the reliability of real-time flows in the network. The main contributions of our work are: (1) A model for estimating end-to-end delay in SDN-based communication networks with non-synchronized clocks is proposed, (2) A fault-resilient contract-based framework with SDN is proposed which aims at managing timing constraints for traffic in time-sensitive industrial operations, (3) A path finding algorithm is presented which is responsible for routing/rerouting real-time flows under different network states.

The proposed framework is aligned with these Industry 4.0 objectives: (1) *self-reconfigurability*: the proposed methodology employs contracts for clearly describing the required guarantees about output of a component. Parallel running observers in the RM are responsible for monitoring violation of a contract (considered as a fault); at the same time, monitors are responsible for detecting events such as link-failure and dynamic changes in the delay requirements. Violation of a contract triggers a suitable response strategy with the help of control logic in order to ensure zero downtime operation. (2) *self-optimization*: the SDN controller attempts to select an optimal path based on current network state in order to achieve maximum quality-of-service (QoS).

From this point forward, the rest of the paper is structured as follows: Section II provides insight to the state-of-the art approaches in this direction. Section III presents the experimental evaluation of the proposed mechanism to estimate end-to-end delay in a network. The details of the proposed contract-based framework are discussed in Section IV. Section V describes our experimental setup and emulation results to show the effectiveness of our proposed approach. Finally, we conclude the paper in Section VI along with future research directions.

II. REVIEW ON SDN-BASED RESILIENCE MECHANISM

In this section, we review the state-of-the art in fault-resilience mechanisms with SDN. In [6], a failure recovery mechanism for SDN is presented which considers flow requirements and assigns back-up rules to forwarding switches. The mechanism builds a backup path by storing a branch which bypasses the failed link. It attempts to speed up the failure recovery time with the use of a global hash table. However, the mechanism suggests to store the backup paths in the SDN switches which have limited memory capacity. In [7], the authors proposed a hybrid (proactive and reactive) two-stage fault tolerant scheme in order to improve resource utilization and link failure recovery latency. A failure recovery module in the proposed scheme is responsible for installing forwarding rules into the switches along with calculating the backup paths for all links as per the QoS requirements. A backup rule generator module installs the backup rules after the detection of a link failure. The proposed scheme can discover the backup paths within limited number of hops. The authors in [8] proposed a scheme which addresses the issues of network congestion and link failures with two controller mechanisms: (1) *fast failover mechanism* deals with pre-establishment of multiple paths for each source-destination pair in the respective switch with the SDN controller, (2) *fast switchover mechanism* deals with monitoring each port of each switch periodically, and switching the flows having lower transmission rate to another route if a port constantly crosses the predefined threshold of the average transmission rate. However, as in [6], the fast failover mechanism raises the issue of storing multiple backup paths in the limited memory of a switch. The authors in [9] proposed an SDN-based resilient communication architecture which aims at satisfying the QoS needs of a microgrid. The SDN controller attempts to solve network issues such as congestion, port down and bandwidth allocation in order to provide resilient services to microgrid. The proposed solution aims at providing delay guarantees, automatic failover and traffic prioritization. However, the work lacks the details of the SDN-based methodology for route migration in order to provide delay guarantees to flows. In [10], the authors presented a framework which addresses end-to-end delay guarantees for time-sensitive flows in hard real-time systems. The solution proposes isolation of different priority flows with separate queues for each flow. It estimates different kinds of delays with static equations, and as a result, it does not estimate these all-important delays in real-time while it is crucial to estimate the delays in real-time for time-critical applications. Furthermore, it presumes the specification of delay requirements of the traffic flows.

In [11], the authors proposed an SDN-based cyber resilience framework for smart grids to protect the substation communication networks against link flood attacks. An SDN controller incorporating a security risk model constantly monitors the network resources. The controller is responsible for detecting heavy flows and congested links to detect link floods. Consequently, the controller assists in selecting an optimal mitigation policy against link flood attacks. At the same time, in order to keep the flows unaffected by the mitigation policies, the controller imposes the QoS policy on all the nodes. The authors in [12] also proposed network resilience strategies with the combination of network functions virtualization (NFV) and SDN against distributed denial-of-service (DDoS) attacks. The authors in [13] view resilience as the restoration of a requisite security state. The work primarily focuses on setting up a framework to attain equilibrium resilience and cybersecurity for SDN-based manufacturing applications. It represents the cyber resilience mechanisms to preserve the required security state in the system along with the operational profiles subject to a disruption event. However, the work does not include implementation of the envisioned resilience framework and does not consider other network failures.

The authors in [14] proposed resilient distribution of controllers for software-defined networks. The proposed scheme aims at maximizing controller distribution to make the network resilient to failures as well as minimizing the flow set-up latency. The proposed controller architecture has global view of the network with a designated controller maintaining a repository of global objects in order to establish control areas by distributed control. In this work, the authors proposed two heuristic mechanisms: (1) to optimize the controller placement to achieve maximum network resilience to avoid network partitioning, (2) to minimize the network area in the case of a node failure. In [15], the authors addressed the research issue of controller placement for improving resilience in the control plane against node failures. Taking controller capacity into consideration, the proposed mechanism aims at maximizing the connectivity with the use of backup controllers adopting definite replication methodology. Each switch contains a list of backup controllers. The proposed mechanism attempts to improve the work presented in [16] which does not take into account the capabilities of controllers and demands of switches. However, the proposed mechanism does not deal with the large scale networks. A network resilience management framework is proposed in [17] which contains policy-controlled management patterns. The framework describes the orchestration of individual resilience services which are distributed over distinct controllers. OpenFlow applications are selected and configured as per the requirements of the specific mechanism for each pattern. The proposed framework is useful for systematic establishment of network-wide resilience services.

In this paper, we propose an approach to address dynamically changing demands of real-time flows unlike the mechanisms proposed in [6], [7], [8], [9], [10]. Additionally, we aim at addressing network resilience for time-sensitive applications without focusing on security aspects as proposed in [11], [12], [13] or, controller placement strategies as proposed in [14], [15], [16], [17]. We focus on the issue of managing delay requirements of time-sensitive flows with a contract-based methodology by incorporating a network resilience mechanism.

## III. END-TO-END DELAY ESTIMATION

It is imperative to accurately estimate end-to-end delay in time-sensitive industrial networks. However, this is a challenging task, especially when the estimation is to be carried out in real-time, because of the: 1) complex topologies, 2) high traffic volume and, 3) requirement of less overhead induced by the estimation mechanism. In this section, we present a delay estimation model, termed *Link-layer Delay Estimator (LLDE)*, and compare the estimated delays with the actual delays.

### A. Design of LLDE

In order to estimate end-to-end delay in real-time for time-critical industrial networks more accurately, it is crucial to consider the transmission delay [18], [19] along with the link delay (propagation delay). For estimation of link delays, LLDE employs the mechanism proposed in [5]. Additionally, LLDE estimates the transmission delays at each of the network switches. Consequently, assuming zero queuing delay, LLDE estimates the end-to-end delay in between two network nodes by aggregating the: 1) estimated link delays and; 2) estimated transmission delays.

Traditional end-to-end delay estimation techniques periodically calculate response times between sent and received probe packets such as *Internet Control Message Protocol (ICMP) Echo* request-reply packets. As a result, these probes are unable to estimate the latency of path segments between arbitrary devices [5]. Other passive estimation techniques [20] require installation of tools or measurement devices on each node which significantly increases cost of deployment for large scale networks. Adding to that, these techniques induce overhead in performing real-time aggregation of measurements captured on individual devices. Software-defined networking can not only provide global network visibility, but can also estimate end-to-end delay in the network with high accuracy, more ease and in an inexpensive way.

*OpenFlow*-based SDN extensively employ *Link Layer Discover Protocol (LLDP)* for discovering and updating network topology. LLDE is a light-weight mechanism based on OpenFlow protocol version 1.3 which works efficiently even when the clocks of the switches are not synchronized. In order to minimize the communication overhead, LLDE utilizes the default SDN packets such as LLDP and Echo packets to discover link delay and transmission delay. The LLDP packet is exploited to store the timestamp information while it is being sent by the SDN controller for topology discovery and updates. Ethernet frames containing LLDP data units are periodically sent by the SDN controller to each of its interfaces. As shown in Figure 2, the controller initiates link discovery procedure by sending a *Packet-Out* message with LLDP at time $t_{cs1}$ which contains *timestamp* ($t_{cs1}$), *Datapath_ID* and *output_port_ID* of switch *S1*. When *S1* receives the LLDP packet, it broadcasts it to all broadcast enabled ports. After the switch *S2* receives the broadcasted packet from *S1*, it notifies the controller using *Packet-In* with LLDP which contains its own *Datapath_ID* and *ingress_port_ID* along with the *timestamp* information ($t_{cs1}$). This is how controller detects a unidirectional link from *S1* to *S2* at time $t_{cs1-s1-s2-c}$. Similarly, the controller detects another unidirectional link (and in this way, the whole network topology) from *S2* to *S1* by sending a *Packet-Out* with LLDP at time $t_{cs2}$ and receiving a *Packet-In* with LLDP at time $t_{cs2-s2-s1-c}$. Meanwhile, OpenFlow Echo messages are sent from the controller to *S1* and *S2* in order to measure the round-trip times (*RTT*), $t_{cs1-s1-c}$ (time to traverse from controller- *S1*-controller) and $t_{cs2-s2-c}$ (time to traverse from controller- *S2*-controller) respectively. As a result, LLDE can estimate the link delay between *S1* and *S2* with equation (1) considering the latency between controller and switches to be half of the RTT:

$$LD_{s1-s2} = ((t_{cs1-s1-s2-c} - t_{cs1}) + (t_{cs2-s2-s1-c} - t_{cs2}) - t_{cs1-s1-c} - t_{cs2-s2-c}) \quad (1)$$

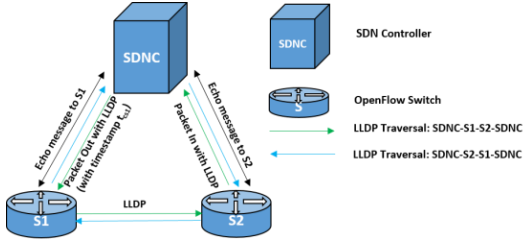

Fig. 2. Link discovery procedure with LLDP

LLDE also emphasizes estimation of the transmission delay, which is periodically measured by packet-length and available bandwidth (calculated by *packet_length/bandwidth*) in real-time while LLDP traverses through the switches. Thus, the transmission delay is calculated at each switch (for *S1* and *S2*, transmission delays are $TD_{s1}$ and $TD_{s2}$ respectively). LLDE, thereby, calculates cost of the link from *S1* to *S2* with equation (2). Additionally, LLDE suggests to store the costs of each link in a graph structure in the SDN controller.

$$LC_{s1-s2} = TD_{s1} + LD_{s1-s2} \quad (2)$$

For a linear topology of *n* switches, the end-to-end delay ($ED_{s1-sn}$) for a packet travelling from switch *S1* to switch *Sn* is estimated with equation (3):

$$ED_{s1-sn} = \sum_{i=1}^{n-1} (LC_{si-s(i+1)}) \quad (3)$$

### B. Performance Evaluation of LLDE

There exist a variety of models to estimate end-to-end delay in a network. As mentioned in the previous subsection, *Ping* (using ICMP packets) is not an efficient method to estimate the path latency, rather it is unfit and unreliable technique to estimate path latency [21]. In this subsection, we evaluate the accuracy of LLDE by comparing its results with the actual end-to-end delays derived by the *iPerf* tool. We implement LLDE in the *Ryu* SDN controller with the *Mininet* emulator. It is to be noted that *iPerf* should not be used to measure end-to-end delay as it generates massive traffic which would interfere with the industrial system traffic.

The experiments are carried out for a network topology of 10 switches (*Sn*=10) with the delay estimation interval of 10 seconds. We consider a linear topology to compare the estimated results (with LLDE) with the actual results as the results can only be compared if there is a single path between the source and the destination. We consider UDP traffic as UDP is widely used for time-sensitive applications.

Each point in a graph is an average of 10 individual results. Table I summarizes the tools used for implementation and experimental evaluation.

TABLE I
TOOLS USED FOR EXPERIMENTS

| Software and Version | Function |
|---|---|
| Ubuntu 16.04 | Host operating system |
| Mininet 2.3.0d4 | Network emulator |
| OpenFlow 1.3 | SDN protocol for southbound interface |
| Ryu 4.26 | SDN Controller |
| Python 2.7.12 | Programming language |
| iPerf | Generating UDP traffic and measuring delay |

*1) Test 1:* In this test, we vary the bandwidth utilization from 0% to 100% for 1 Mbps links with suitable UDP flows from the source (*S1*) to the destination (*S10*).

As shown in Figure 3, as bandwidth utilization increases from 0% to 100%, estimated delays by LLDE tend to increase due to increase in the network traffic. While LLDE estimates the average delay of 0.037 ms (average of results for all four bandwidth utilizations), the actual average end-to-end delay measured with iPerf is 0.039 ms.

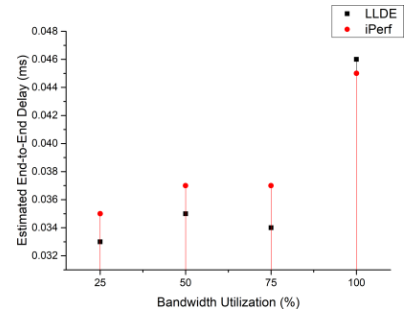

Fig. 3. Delay estimation comparison with 1 Mbps link bandwidth

*2) Test 2:* In this test, we vary the bandwidth utilization from 0% to 100% for 100 Mbps links with suitable UDP flows from the source to the destination.

As shown in Figure 4, while LLDE estimates the average delay of 0.070 ms, the actual average end-to-end delay measured with iPerf is 0.063 ms.

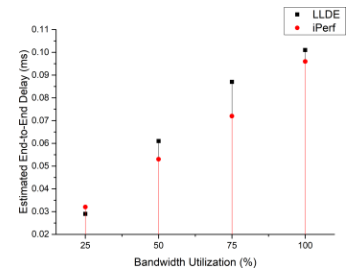

Fig. 4. Delay estimation comparison with 100 Mbps link bandwidth

*3) Test 3:* In this test, we vary the bandwidth utilization from 0% to 100% for 1 Gbps links with suitable UDP flows from the source to the destination.

As shown in Figure 5, while LLDE estimates the average delay of 0.060 ms, the actual average end-to-end delay measured with iPerf is 0.058 ms.

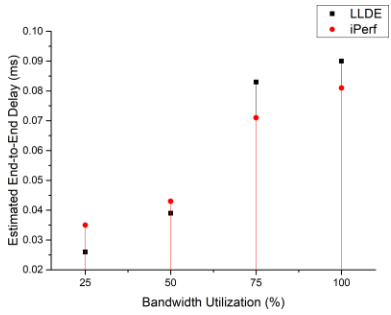

Fig. 5. Delay estimation comparison with 1 Gbps link bandwidth

From the experiments, we can see that LLDE provides accuracy in end-to-end delay estimation in tens of microseconds. Therefore, the path finding algorithm in the proposed resilience management framework employs LLDE for the estimation of end-to-end delay.

## IV. PROPOSED RESILIENCE MANAGEMENT FRAMEWORK

In this section, we present an SDN-based network resilience management framework, termed as *SDN-RM*, which employs a contract-based methodology that aims at satisfying end-to-end delay requirements of all the messages of a flow. The *end-to-end delay requirement* is defined as the need of a message to travel from the source to the destination within the specified time period, while a *flow* is defined as network traffic with a series of packets from a source to a destination.

Figure 6 represents the architectural design of the resilience management framework residing in the SDN controller. It is to be noted that the SDN controller can be placed in a server which has high computation power.

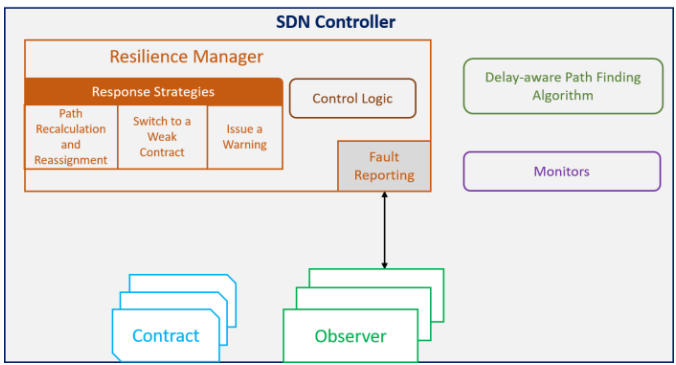

Fig. 6. Architectural design of the resilience management framework

### A. Contracts and Observers

A *contract* [22] is employed to detect faults more robustly and quickly. In this framework, a contract precisely describes: (1) inputs and outputs of a component, (2) assumptions on the inputs and environment, (3) parameters which could be used for run-time updates to the contracts, and (4) guarantees about the outputs of the component. We consider two types of contracts: strong contract and weak contract. Following (Figure 7) is an example of a strong contract which we define for providing end-to-end delay guarantees from a source $si$ to a destination $sj$. This contract states guarantee of the estimated end-to-end delay (estimated during delay estimation interval), termed as $ED_{si-sj}$, to be less than or equal to the end-to-end delay requirement of the flow from $si$ to $sj$, termed as $PED_{si-sj}$. We make the following assumption about the network: (1) Existence of a stable path from $si$ to $sj$ between the delay estimation intervals; (2) Non-occurrence of the two events (mentioned in the following subsection) between the delay estimation intervals.

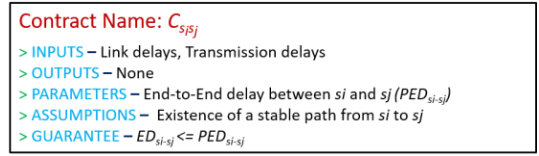

Fig. 7. A sample contract

The corresponding weak contract is defined in the similar way with a weaker set of parameter values. In the case of a fault, the component may switch to the corresponding weak contract in order to achieve zero downtime production. In this work, the contracts were generated manually based on user-provided requirements.

An *observer* for a given contract is responsible to verify whether the expected behavior is generated by the component. In the case of a contract failure (termed as a *fault*) detection by the corresponding observer, a fault is reported. An observer is independent of the component's behavior.

### B. Monitors

Monitors are employed to detect two events: (*E1*) link failure, and (*E2*) run-time updates in the delay requirements (as these events may induce faults).

- *E1*. A link failure monitor in the SDN controller detects link breaks with the received switch port statistics.
- *E2*. A contract modification monitor detects run-time changes in the delay requirements by observing modifications in contracts.

### C. Resilience Manager

The *resilience manager* (*RM*) aims at providing network resilience to the faults. It contains a control logic which receives the reported fault, and accordingly it executes a desired response strategy. The response strategies can be:

- *RS1*. Path recalculation and path reassignment: SDN not only provides global view of the network, but also provides flexible network controlling. Thereby, the RM residing in the SDN controller provides an efficient way for path restoration. Figure 8 represents the path restoration in an SDN-based ICPS by considering a contract failure due to link breakage (E1) between two switches. The path finding algorithm is executed to recalculate the path between the desired source-destination pair in order to reroute the real-time flows. The revised forwarding rules for the alternative paths are deployed by the SDN controller onto the respective switches. Consequently, path restoration delay [23] is calculated (refer Figure 9) as the sum of delays incurred during fault detection, path recalculation and path reassignment.

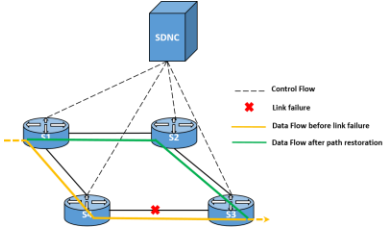

Fig. 8. Path restoration in SDN-based networks

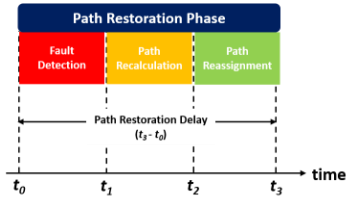

Fig. 9. Path restoration phase and corresponding delay

- *RS2*. Switching to a weak contract: The RM may decide to switch to a weak contract if the delay requirements as per the strong contract cannot be guaranteed after path recalculation.
- *RS3*. Issuing a warning: In the worst case, when the delay requirements of the weak contract cannot be satisfied (weak contract failure), a warning is issued about the same.

### D. Delay-aware Path Finding Algorithm

The path finding algorithm is responsible to find the best path in terms of end-to-end delay. It is triggered: (1) when a new flow arrives, (2) periodically to suggest a new path based on the current network state, and (3) in case of a fault is reported by the observer. The SDN controller then adopts the newly suggested path for the on-going flows. Considering an SDN-based network as a graph $G=(S,L)$, where $S$ represents a set of switches and $L$ represents a set of links between the switches (subject to failure), the cost of a link is calculated by LLDE (refer Section III). The path-finding algorithm employs *Dijkstra's shortest path algorithm* to discover a path with shortest delay between the source ($Ssrc$) and the destination ($Sdst$). The inputs to the algorithm are $G$, $Ssrc$, $Sdst$ and cost matrix stored in the controller. Output of the algorithm is an optimal path ($P_{Ssrc-Sdst}$) from $Ssrc$ to $Sdst$ and end-to-end delay ($ED_{ssrc-sdst}$) of $P_{Ssrc-Sdst}$.

## V. RESULTS AND ANALYSIS

In this section, we evaluate the performance of SDN-RM by answering the following questions:

- *Q1*. What is the success rate (resilience) of SDN-RM in the case of faults caused by link failure (E1) and dynamic change in delay requirements (E2)?
- *Q2*. What is the throughput provided by SDN-RM in this case?
- *Q3*. How much is the path restoration delay of SDN-RM in this case?

The performance of our proposed framework, SDN-RM, is compared with: (1) SDN controller without any resilience management, termed as *SDN-woRM*, (2) SDN controller having resilience management with strong contracts (without considering weak contracts), termed as *SDN-sRM*, (3) SDN controller having resilience management with only proactive strategy which detects fault only after a specific time period (during the delay estimation interval), termed as *SDN-pRM*. Table II summarizes the characteristic comparison of the four mechanisms:

TABLE II
CHARACTERISTIC COMPARISON BETWEEN DIFFERENT MECHANISMS

| Mechanism | Proactive Strategy | Reactive Strategy | Strong Contracts | Weak Contracts |
|---|---|---|---|---|
| SDN-woRM | x | x | x | x |
| SDN-sRM | ✓ | ✓ | ✓ | x |
| SDN-pRM | ✓ | x | ✓ | ✓ |
| SDN-RM | ✓ | ✓ | ✓ | ✓ |

The performance of the aforementioned mechanisms is evaluated with these three metrics: *success rate* (determined by number of times the delay requirements of flows are satisfied), *throughput* and *path restoration delay*. The experiments are carried out on different topologies with Mininet testbed. Table III summarizes the experimental parameters. In the experiments, the delay estimation interval is set to 10 seconds as it is reasonable to assume a stable network state during this time period [5].

TABLE III
EMULATION PARAMETERS

| Parameters | Value |
|---|---|
| Number of switches | 10 |
| Number of hosts | 8 |
| Number of flows (Varying) | 2 to 10 |
| Link capacity | 1 Gbps |
| Emulation time | 150 seconds |
| Delay estimation interval | 10 s |
| Number of Events E1or/and E2 (Varying) | 1 to 5 |
| Traffic type | UDP |
| Traffic generation | iPerf |

### A. Tests

In the following tests, we consider the experimental topology as shown in Figure 10.

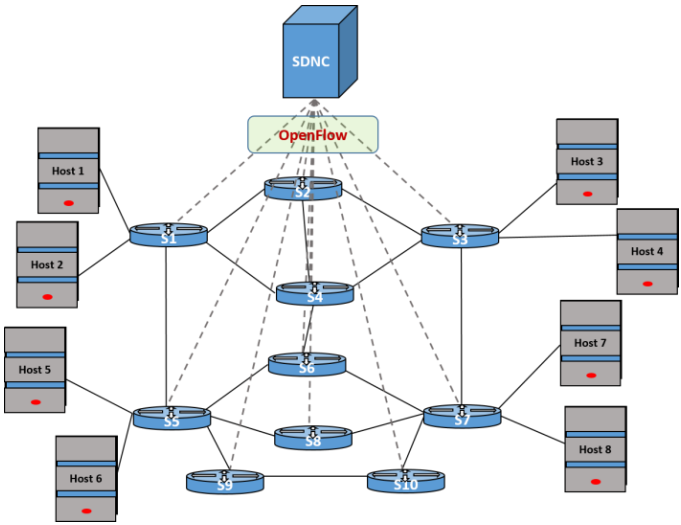

Fig. 10. Experimental testbed

*1) Varying Number of Flows:* In this test, we vary the number of parallel flows (utilizing higher percentage of bandwidth) from 2 to 10 for a source-destination pair (*Host 1* to *Host 8*) by triggering events (E1, E2) 4 times in order to generate multiple faults during the emulation period. We consider sending each flow of 100 Mb with a time interval of 1 second.

*a) Test 1: Regarding link failure:* During the emulation, the link is broken (E1) 4 times between distinct switches in the presented industrial network.

As shown in the Figure 11, as the number of flows increases, the success rate of all the mechanisms decreases due to increased network congestion. It is obvious that the average success rate of SDN-woRM is 56.40% as it does not have any resilience mechanism. While SDN-sRM provides the average success rate of 70.53%, SDN-pRM improves it by nearly 3%. This suggests that it is important to plan the worst-case flow requirements in the case of faults. Meanwhile, SDN-RM provides the average success rate of 81.47% as it not only possesses proactive and reactive strategies (unlike SDN-pRM), but it also contains pre-determined worst-case flow requirements of the flows (unlike SDN-sRM). Thus, SDN-RM provides improved network resilience.

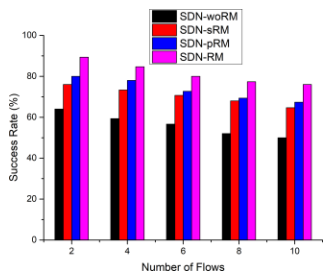

Fig. 11. Success rate during link-breaks (E1) with varying number of flows

As shown in the Figure 12, it is obvious that the network throughput increases as the number of flows increases. As SDN-woRM does not contain any resilience mechanism, it provides an average network throughput of 404.2 Mbps. Meanwhile, due to aforementioned reasons, SDN-sRM and SDN-pRM provide an average throughput of 579.8 Mbps and 587.6 Mbps respectively, while SDN-RM notably improves the network QoS by providing an average throughput of 592.6 Mbps.

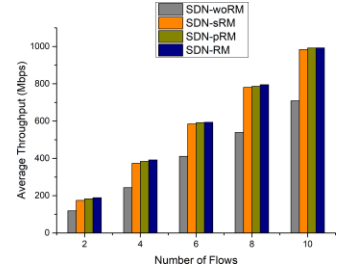

Fig. 12. Throughput during link-breaks (E1) with varying number of flows

It is to be noted that the average path restoration delay for SDN-RM is 0.431 ms, while that of SDN-sRM is 0.300 ms. At the same time, due to lack of any reactive strategy SDN-pRM provides the average restoration delay of 10.076 seconds.

*b) Test 2: Regarding run-time changes in delay requirements:* In this test, the delay requirements have been changed (E2) at 4 different time instances during the execution. The results of Figure 13 and Figure 14 are summarized in Table IV.

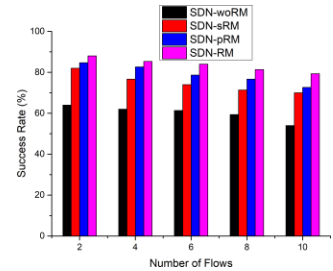

Fig. 13. Success rate during dynamic change in requirements (E2) with varying number of flows

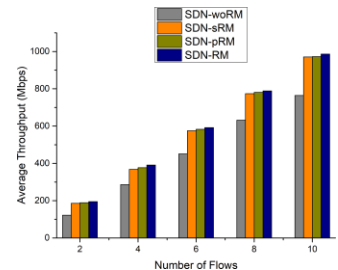

Fig. 14. Throughput during dynamic change in requirements (E2) with varying number of flows

*c) Test 3: Regarding link failure and run-time changes in delay requirements:* During the emulation, the link is broken

TABLE IV
TEST 2: PERFORMANCE EVALUATION

| Mechanism | Success Rate | Throughput | Path Restoration Delay |
|---|---|---|---|
| SDN-woRM | 60.13 % | 451 Mbps | - |
| SDN-sRM | 74.80 % | 574.8 Mbps | 0.359 ms |
| SDN-pRM | 79.07 % | 580.6 Mbps | 10.150 seconds |
| SDN-RM | 83.60 % | 590.6 Mbps | 0.539 ms |

(E1) twice between distinct switches in the presented industrial network; moreover, the delay requirements have been changed (E2) at 2 different time instances during the execution. The results of Figure 15 and Figure 16 are summarized in Table V.

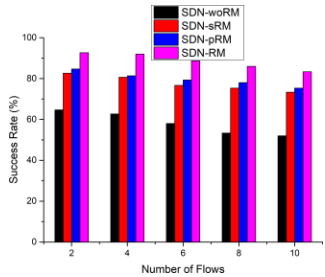

Fig. 15. Success rate during events (E1 and E2) with varying number of flows

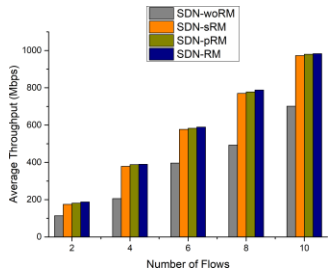

Fig. 16. Throughput during events (E1 and E2) with varying number of flows

TABLE V
TEST 3: PERFORMANCE EVALUATION

| Mechanism | Success Rate | Throughput | Path Restoration Delay |
|---|---|---|---|
| SDN-woRM | 58.13 % | 382 Mbps | - |
| SDN-sRM | 77.73 % | 574.8 Mbps | 0.293 ms |
| SDN-pRM | 79.73 % | 582 Mbps | 10.117 seconds |
| SDN-RM | 88.53 % | 587.6 Mbps | 0.434 ms |

*2) Varying Number of Events:* In order to evaluate SDN-RM more rigorously, in this test, we vary the number of events (E1 and E2) from 1 to 5 by keeping number of flows to 6. The other parameters are set same as above three tests.

  *a) Test 4: Regarding link failure:* During the emulation, the link is broken (E1) from 1 to 5 times between distinct switches in the presented industrial network. As shown in the Figure 17, as the number of link breaks increases, the success rate of all the mechanisms decreases due to increased number of contract violations (faults). It is obvious that the average success rate of SDN-woRM is 61.73% as it does not have any resilience mechanism. While SDN-sRM provides the average success rate of 77.33%, SDN-pRM improves it by nearly 2%. Meanwhile, SDN-RM provides the average success rate of 85.47% due to the aforementioned reasons. Thus, SDN-RM provides improved network resilience even in the case when multiple faults are generated.

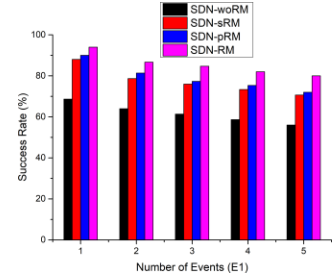

Fig. 17. Success rate with varying number of link breaks (E1)

As shown in the Figure 18, it is obvious that the network throughput decreases as the number of link failure increases. As SDN-woRM does not contain any resilience mechanism, it provides an average network throughput of 476.6 Mbps. Meanwhile, SDN-sRM and SDN-pRM provide an average throughput of 573.4 Mbps and 578.8 Mbps respectively, while SDN-RM considerably improves the network QoS by providing an average throughput of 584.2 Mbps.

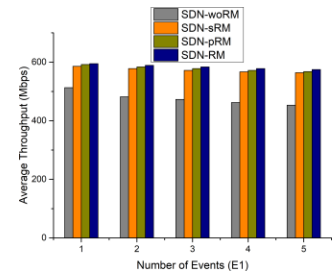

Fig. 18. Throughput with varying number of link breaks (E1)

The average path restoration delays for SDN-RM, SDN-sRM and SDN-pRM are 0.539 ms, 0.359 ms and 10.175 seconds.

  *b) Test 5: Regarding run-time changes in delay requirements:* In this test, the delay requirements have been changed (E2) from 1 to 5 times at different time instances during the emulation. The results of Figure 19 and Figure 20 are summarized in Table VI.

  *c) Test 6: Regarding link failure and run-time changes in delay requirements:* During the emulation, both the events (E1 and E2) occur from 1 to 5 times at different time instances during the execution. The results of Figure 21 and Figure 22 are summarized in Table VII.

The average path restoration delays for SDN-RM, SDN-sRM and SDN-pRM are ms, ms and 10.135 seconds.

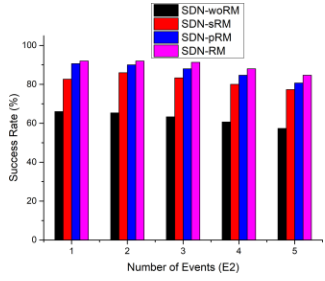

Fig. 19. Success rate with varying number of changes in delay requirements (E2)

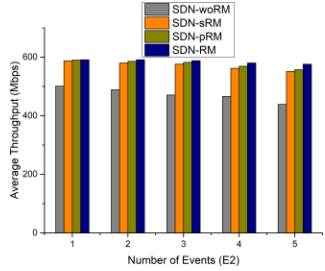

Fig. 20. Throughput with varying number of changes in delay requirements (E2)

*Remark* 1. The tests suggest that the planning of the worst-case delay requirements (with mechanisms such as weak contracts) is vital. At the same time, resilience framework having only proactive strategy not only increases the path restoration delay, but also degrades QoS in the network (it is to be noted that performance of SDN-pRM depends upon the selected time interval). We note that it is imperative to have both, proactive and reactive strategies. SDN-RM provides improved performance over all the other mechanisms as it takes all four aspects into considerations (as shown in Table

TABLE VI
TEST 5: PERFORMANCE EVALUATION

| Mechanism | Success Rate | Throughput | Path Restoration Delay |
|---|---|---|---|
| SDN-woRM | 62.53 % | 473.2 Mbps | - |
| SDN-sRM | 81.87 % | 571.2 Mbps | 0.327 ms |
| SDN-pRM | 86.80 % | 576.8 Mbps | 10.149 seconds |
| SDN-RM | 89.60 % | 585.2 Mbps | 0.308 ms |

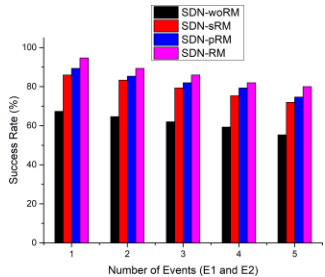

Fig. 21. Success rate with varying number of events (E1 and E2)

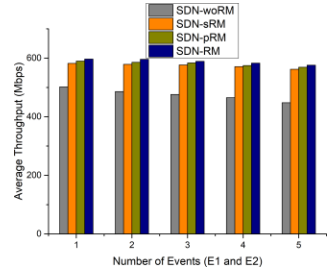

Fig. 22. Throughput with varying number of events (E1 and E2)

TABLE VII
TEST 6: PERFORMANCE EVALUATION

| Mechanism | Success Rate | Throughput | Path Restoration Delay |
|---|---|---|---|
| SDN-woRM | 61.73 % | 475.4 Mbps | - |
| SDN-sRM | 79.20 % | 574.2 Mbps | 0.311 ms |
| SDN-pRM | 82.13 % | 580.8 Mbps | 10.135 seconds |
| SDN-RM | 86.40 % | 588.2 Mbps | 0.327 ms |

II).

### B. Tests on Different Settings

In the following tests, we change the experimental settings to verify the remark drawn from the previous tests. We consider a mesh topology of 20 switches with a host connected to each switch as shown in Figure 23. We set the number of parallel flows to 3 (with each flow of 200 Mb generated after a time interval of 2 seconds) for five different source-destination pairs. We vary the number of events (E1 for the whole network and, E2 for each source-destination pair) from 1 to 5 to generate multiple faults during the emulation period.

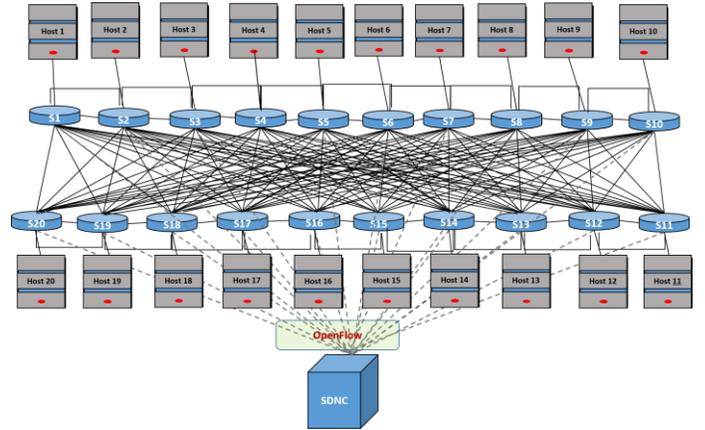

Fig. 23. Another experimental testbed

*a) Test 7: Regarding link failure:* During the emulation, the link is broken (E1) from 1 to 5 times between distinct switches in the network. As shown in the Figure 24, as the number of link breaks increases, the success rate of all the mechanisms decreases due to increased number of faults.

The average success rate of SDN-woRM is 59.76% as it does not have any resilience mechanism. While SDN-

sRM provides the average success rate of 78.93%, SDN-pRM improves it by nearly 3%. Meanwhile, SDN-RM provides the average success rate of 86.16% due to the aforementioned reasons. Thus, SDN-RM provides improved network resilience even under different network settings.

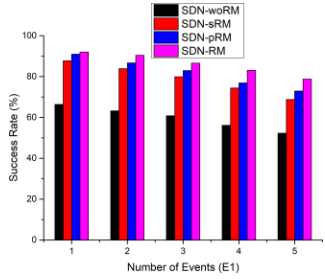

Fig. 24. Success rate with varying number of link breaks (E1)

As shown in the Figure 25, the network throughput decreases as the number of link failure increases. SDN-woRM provides an average network throughput of 464.16 Mbps. Meanwhile, SDN-sRM and SDN-pRM provide an average throughput of 567.36 Mbps and 571.16 Mbps respectively, while SDN-RM considerably improves the network QoS by providing an average throughput of 577.56 Mbps.

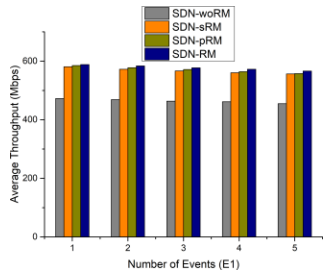

Fig. 25. Throughput with varying number of link breaks (E1)

The average path restoration delays for SDN-RM, SDN-sRM and SDN-pRM are 0.424 ms, 0.430 ms and 10.303 seconds.

*b) Test 8: Regarding run-time changes in delay requirements:* In this test, the delay requirements have been changed (E2) from 1 to 5 times for each pair at different time instances during the emulation. The results of Figure 26 and Figure 27 are summarized in Table VI.

TABLE VIII
TEST 8: PERFORMANCE EVALUATION

| Mechanism | Success Rate | Throughput | Path Restoration Delay |
|---|---|---|---|
| SDN-woRM | 60.51 % | 467.4 Mbps | - |
| SDN-sRM | 79.12 % | 563.2 Mbps | 0.338 ms |
| SDN-pRM | 82.40 % | 571 Mbps | 9.951 seconds |
| SDN-RM | 87.25 % | 578.9 Mbps | 0.461 ms |

*c) Test 9: Regarding link failure and run-time changes in delay requirements:* During the emulation, both the events

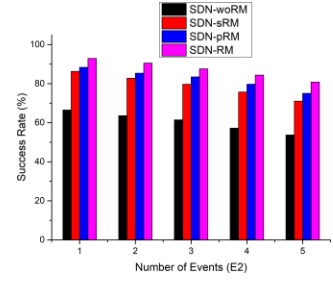

Fig. 26. Success rate with varying number of changes in delay requirements (E2)

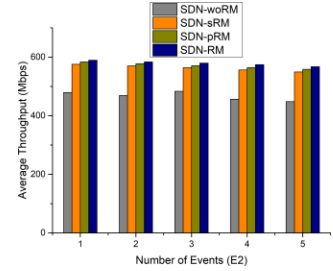

Fig. 27. Throughput with varying number of changes in delay requirements (E2)

(E1 for the whole network and, E2 for each pair) occur from 1 to 5 times at different time instances during the execution. The results of Figure 28 and Figure 29 are summarized in Table IX.

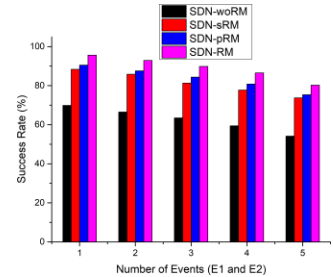

Fig. 28. Success rate with varying number of events (E1 and E2)

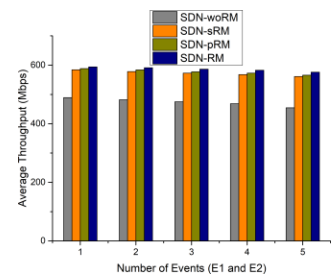

Fig. 29. Throughput with varying number of events (E1 and E2)

*Remark* 2. The tests on the different network settings confirm the *Remark 1*. Moreover, the tests show that the proposed

TABLE IX
TEST 9: PERFORMANCE EVALUATION

| Mechanism | Success Rate | Throughput | Path Restoration Delay |
|---|---|---|---|
| SDN-woRM | 62.69 % | 473.8 Mbps | - |
| SDN-sRM | 81.41 % | 572.8 Mbps | 0.411 ms |
| SDN-pRM | 83.73 % | 577.6 Mbps | 10.300 seconds |
| SDN-RM | 89.04 % | 586 Mbps | 0.423 ms |

framework provides scalability.

## VI. CONCLUSION

In this paper, firstly, we present a mechanism to estimate end-to-end delay in an SDN-based network and present the experimental proofs to depict the accuracy provided by the mechanism in estimating end-to-end delay. Then we propose a contract-based network resilience management framework, termed SDN-RM, for critical real-time flows in industrial cyber-physical systems. Delay-based contracts are used to define guarantees about the component's behavior and violation of a contract is monitored using observers. The proposed framework detects distinct events (with monitors), detects faults (contract violation) and rapidly recover from multiple faults by discovering an alternate path (with a delay-aware path finding algorithm). The proposed framework is evaluated with three important metrics: success rate of satisfying delay requirements, network throughput and path restoration delay. Extensive emulations under distinct network conditions demonstrate that SDN-RM provides higher success rate and improved network throughput with notable path restoration delay (in fraction of milliseconds). Consequently, the results suggest that it is imperative for a resilience manager to have a rapid reaction mechanism and worst-case delay requirements for time-critical industrial applications.

In future, we plan to develop a bandwidth and delay aware routing algorithm which can balance traffic load in the industrial networks. Furthermore, we plan to consider priority of flows in order to meet their QoS requirements as per the determined level of criticality.

## ACKNOWLEDGEMENT

This research work was conducted within the Delta-NTU Corporate Laboratory for Cyber-Physical Systems with funding support from Delta Electronics Inc. and the National Research Foundation (NRF), Singapore under the Corp Lab @ University Scheme. We would like to acknowledge the contributions from Sidharta Andalam and Jeffrey Soon of Delta Electronics, Singapore for their valuable inputs.